\documentclass[aps,pra,nofootinbib,twocolumn]{revtex4-2}

\usepackage{graphicx,amsmath,amssymb}

\usepackage{dcolumn}

\usepackage{bm}

\begin{document}

\title{Direct estimation of $g^{(2)}(0)$ from click statistics of a multiplexed detector}

\author{Jaromír Fiurášek}

\email[Corresponding author: ]{fiurasek@optics.upol.cz}
\affiliation{Department of Optics, Faculty of Science, Palack\'y University, 17.\ listopadu 12, 77900  Olomouc, Czech Republic}

\begin{abstract}
We present and theoretically analyze a method for direct estimation of the second-order correlation function $g^{(2)}(0)$ from the click statistics of a multiplexed detector composed of $M$ binary on-off detection channels. The method uses linear estimators of the first and second moments of the photon-number operator and can provide accurate results even for optical fields with substantial mean photon numbers. For a fixed number of channels, the detection range can be extended by introducing additional  losses before the detector. We demonstrate the performance of the method for coherent, thermal, amplitude-squeezed, and photon-added coherent states. We also derive a nonclassicality criterion based on the estimated correlation function and show how the method can be adapted to unbalanced detectors.
Our work establishes direct estimation of \(g^{(2)}(0)\) as a simple and versatile method for quantitatively characterizing quantum light using multiplexed click detectors.
\end{abstract}

\maketitle

\section{Introduction}

Experimental characterization of  photon number distributions of various quantum light sources and signals plays a fundamental role in  quantum  optics and optical quantum technologies.  Since the determination of the full photon number distribution $p_n$  is often challenging, one typically resorts to measurements of some key characteristics of $p_n$. A widely used quantity in this context is the normalized second-order auto-correlation function $g^{(2)}(\tau)$ evaluated at $\tau=0$  \cite{Loudon2000},
\begin{equation}
g^{(2)}(0)=\frac{\langle \hat{a}^{\dagger 2}\hat{a}^2\rangle }{\langle \hat{a}^\dagger \hat {a}\rangle^2}= \frac{\langle \hat{n}(\hat{n}-1)\rangle }{\langle \hat{n}\rangle^2}.
\label{g2definition}
\end{equation}
Here $\hat{a}$ and $\hat{a}^\dagger$ denote the annihilation and creation operator, respectively, and $\hat{n}=\hat{a}^\dagger \hat{a}$ is the photon number operator.
Since $g^{(2)}(0)$ is defined in terms of the first and second normally ordered moments of the photon number operator, it carries information on the relationship between the mean photon number and the photon number variance. In particular,  $g^{(2)}(0)=1$ for coherent light exhibiting Poissonian photon number statistics, and $g^{(2)}(0)<1$ indicates that the light is nonclassical and possesses a sub-Poissonian photon number distribution.

\begin{figure}[b]
\includegraphics[width=\linewidth]{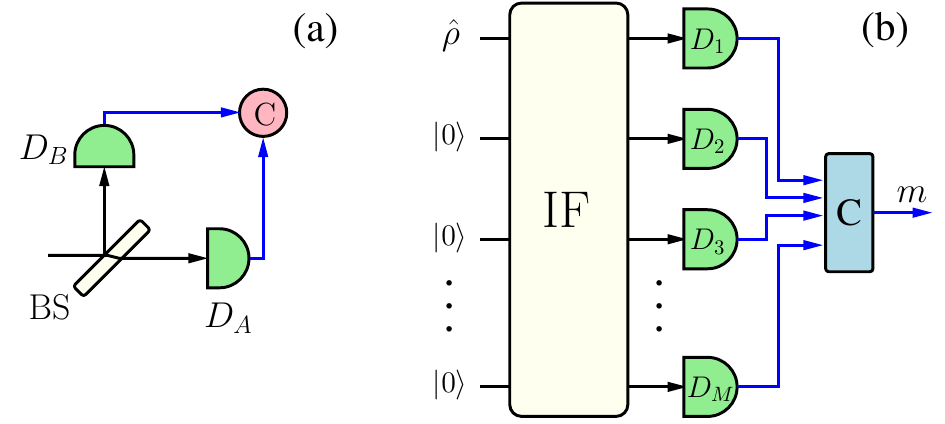}
\caption{(a) The Hanbury Brown and Twiss detection scheme. The input signal is split at a balanced beam splitter BS and measured with two click detectors $D_A$ and $D_B$. Coincidences of detector clicks are counted, together with single clicks.  (b) Multiplexed $M$-channel detector. The input signal is divided into $M$ spatial or temporal channels using an $M$-port interferometer. Each output mode is measured with a click detector that distinguishes the presence and absence of photons, and the number of clicks is counted.}
\label{g2fig1}
\end{figure}

Measurements of the second-order correlation function $g^{(2)}(\tau)$ have a long history dating back to the pioneering experiments by Hanbury Brown and Twiss (HBT) \cite{HBT1956}. The HBT setup, conceptually illustrated in Fig.~\ref{g2fig1}(a),  has subsequently been employed in countless experiments to characterize sources ranging from single-photon emitters to coherent and chaotic light. In the HBT setup, the input optical signal is split into two parts at a balanced beam splitter, each output mode is  measured with an optical detector, and the correlations between the two detections are recorded. When characterizing faint quantum light,  binary on-off detectors that can only distinguish the presence and absence of photons are typically employed. Each detector in the setup then responds either with a click or no-click and  $g^{(2)}(0)$  is usually estimated as the ratio of the probability of coincidence clicks $c_{AB}$ and the product of probabilities $s_{A}$ and $s_B$  of clicks of the single detectors  $D_A$ and $D_B$, 
\begin{equation}
g^{(2)}(0) \approx \frac{c_{AB}}{s_A s_B}.
\end{equation}
It is well understood that this relation is only approximate and the quantity estimated in this way approaches the true value of $g^{(2)}(0)$ only in the limit of extremely weak signal \cite{Stensson2018}.

Current quantum optical detection schemes based on the click detectors often go significantly beyond the elementary HBT setup and they employ spatial  or temporal multiplexing to increase the number of detection channels \cite{Paul1996,Rehacek2003,Banaszek2003,Achilles2003,Fitch2003,Micuda2008,Kalashnikov2011,Bartley2013,Mattioli2016,Moreva2017,Straka2018,Kroger2017,Sperling2017a,Sperling2017b,Tiedau2019,
Hlousek2019,Lachman2019,Tiedau2021,Cheng2023,Knoll2003,Santana2024,Sullivan2024}. 
This increases the  information gained by the measurement, enables measurement of  higher-order correlation functions \cite{Avenhaus2010,Kalashnikov2011,Laiho2012,Guo2020}, and enables characterization of correlations in twin beams \cite{Allevi2012,Sperling2013,Sperling2015,Sperling2016}. One can even attempt to reconstruct the photon number distribution from the measured click statistics \cite{Hlousek2019,Banner2024,Matsumoto2026}, and then determine $g^{(2)}(0)$ from the reconstructed $p_n$ \cite{Hlousek2019}. However, the click statistics generally contains only partial information on the photon number distribution, imposing fundamental limits on the ability to reconstruct $p_n$ from the click statistics \cite{Kumazawa2019,Fiurasek2025}.

Here we investigate a method to estimate the second-order correlation function directly from the click statistics. Specifically, we propose to utilize carefully designed linear estimators of the first and second moments of the photon number operator, $\langle \hat{n}\rangle$ and $\langle \hat{n}^2\rangle$. 
These estimators are constructed such that they work exactly for states that contain no more than $M$ photons, where $M$ is the number of detection channels. 
Our approach is thus closely related to the approximate reconstruction of photon number distribution  via  linear pseudoinversion of the relation between click statistics and $p_n$ \cite{Kovalenko2018,Ares2024}. 
As we show below, we can achieve very precise estimation of $g^{(2)}(0)$  even for fields with mean photon numbers significantly exceeding $1$. The estimated $g_M^{(2)}(0)$ can be treated as a specific nonclassicality witness and we establish an exact nonclassicality criterion for it that we prove for balanced detectors.

The rest of the paper is organized as follows. In Section II we review the theoretical description of the multiplexed detectors and the relation between the measured click statistics and the photon number distribution. The proposed method for direct estimation of $g^{(2)}(0)$ from the click statistics is presented in Section III. An exact  nonclassicality criterion for the estimated correlation function is presented  in Section IV. In Section V we consider unbalanced detectors and address the impact of different overall detection efficiencies for each detection channel.
Finally, Section VI contains a brief summary and conclusions.

\section{Multiplexed detector }
A spatially multiplexed detector with $M$ detection channels is illustrated in  Fig.~1(b). The input signal is split into $M$ modes using a suitably designed $M$-port optical interferometer. Each output mode is measured with a binary on-off detector that can distinguish the presence and absence of photons, and produces either a click or no click.  The click statistics, defined as the probability that $m$ detectors click and $M-m$ detectors do not click, is related to the photon number distribution $p_n$ of the input state $\hat{\rho}$ via
\begin{equation}
c_m=\sum_{n=0}^\infty C_{mn} p_n.
\label{cmsum}
\end{equation}
Here $C_{mn}$ describes the response of the multiplexed detector.  Assuming a perfectly balanced $M$-channel detector with overall detection efficiency $\eta$ and negligible dark counts, we get \cite{Hlousek2019}
\begin{equation}
C_{mn}= {M\choose  m}\sum_{j=0}^m { m \choose j} (-1)^j \left[1-\eta +\frac{\eta(m-j)}{M}\right]^n.
\label{Cmndefinition}
\end{equation}
For $\eta=1$ the sum in Eq. (\ref{Cmndefinition}) can be expressed in terms of the Stirling numbers of the second kind \cite{Miatto2018}.
It holds that $C_{mn}=0$ for $m>n$ identically, because at least $m$ photons are required to produce the $m$-fold coincidence. Therefore, the summation in Eq. (\ref{cmsum}) in fact runs from $n=m$ to infinity. 

This changes when dark counts are taken into account because in this case some detectors can be triggered by noise instead of the input signal photons. Assuming a simple model where the probability of dark count $R_D$ is the same for each detector and the dark counts are statistically independent \cite{Sperling2012b}, we obtain
\begin{eqnarray}
C_{mn}&=& {M\choose  m}\sum_{j=0}^m { m \choose j} (-1)^j (1-R_D)^{M-m+j} \nonumber \\
& & \times \left[1-\eta +\frac{\eta(m-j)}{M}\right]^n,
\label{Cmndarkcounts}
\end{eqnarray}
which is a straightforward generalization of Eq. (\ref{Cmndefinition}).

In the experiments, the coincidence time window is typically set to a few nanoseconds while the dark count rates are of the order of $100$ counts per second or smaller. For such configurations  $R_D$ is of the order of $10^{-6}$ . We therefore neglect dark counts in the following calculations and set $R_D=0$. Moreover, the parameter $g^{(2)}(0)$  is insensitive to losses. We can therefore set $\eta=1$ when constructing the estimators.  We emphasize that the estimation procedure developed below is universal and can in principle be applied to an arbitrary multiplexed detector with known response $C_{mn}$. We make use of this fact later on in Section V where we discuss the effect of imbalances of the detector channels.

\section{Estimation of $g^{(2)}(0)$}

As shown in  Refs. \cite{Kovalenko2018,Ares2024}, it is possible to approximately reconstruct the photon number statistics from the click statistics by a simple linear pseudoinversion on a subspace spanned by the $M+1$  lowest  Fock states up to $|M\rangle$. Here, we utilize this approach to explicitly construct
  linear estimators  $\mu_1$ and $\mu_2$ for the first and second moments of the photon number operator $\langle \hat{n}\rangle$ and $\langle \hat{n}^2\rangle$. 
  Subsequently, the estimated values of $\langle \hat{n}\rangle$ and $\langle \hat{n}^2\rangle$ are used to estimate $g^{(2)}(0)$.
 The linear estimators $\mu_1$ and $\mu_2$ are defined as weighted linear combinations of the click probabilities $c_m$ \cite{Fiurasek2025},
\begin{equation}
\mu_1=\sum_{m=0}^M D_{m}c_m, \qquad \mu_2=\sum_{m=0}^M E_m c_m.
\label{mu12definition}
\end{equation}
When processing experimental data, the probabilities $c_m$ are approximated by the measured relative frequencies of clicks of $m$ detectors.

\begin{figure}
\includegraphics[width=\linewidth]{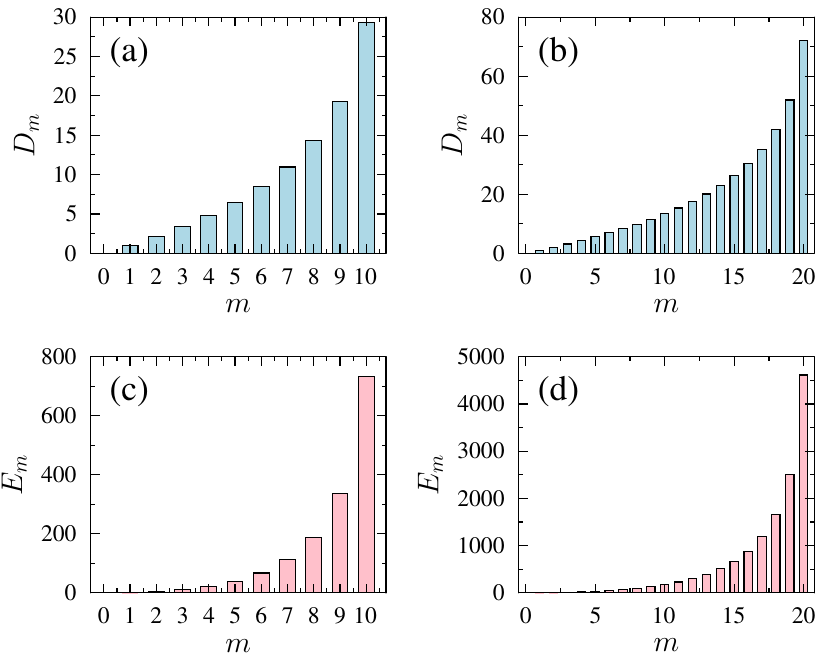}
\caption{The coefficients $D_m$ (a,b) and $E_{m}$ (c,d) that specify the estimators $\mu_1$ and $\mu_2$ are plotted for two multiplexed detectors with different numbers of detection channels  $M=10$  (a,c) and $M=20$ (b,d) \cite{Fiurasek2025}. The estimators are constructed for perfectly balanced detectors with $\eta=1$ and $R_D=0$.}
\label{figDE}
\end{figure}

The estimators (\ref{mu12definition}) are constructed to provide unbiased estimation of $\langle \hat{n}\rangle$ and $\langle \hat{n}^2\rangle$ for any state that contains no more than $M$ photons \cite{Ares2024,Fiurasek2025}. This is achieved when 
\begin{equation}
\sum_{m=0}^M D_m C_{mn}=n, \qquad  \sum_{m=0}^M E_m C_{mn}=n^2, \qquad n\leq M.
\end{equation}
This is a system of linear equations for $D_m$ and $E_m$ which can easily be solved. To proceed, 
 we define finite square matrix $B_{mn}=C_{mn}$, $0 \leq m,n\leq M$. We also define two row vectors $d$ and $e$ with components
\begin{equation}
d_n=n, \qquad e_n=n^2, \qquad 0 \leq  n\leq M.
\end{equation}
The vectors $D$ and $E$ that specify the two estimators (\ref{mu12definition}) can be expressed as
\begin{equation}
D=d B^{-1}, \qquad E=eB^{-1}.
\end{equation}
For $\eta=1$, the elements of the inverse matrix $B^{-1}$ can be expressed analytically in terms of the signed Stirling numbers of the first kind $s(m,n)$ \cite{Ares2024}. We thus obtain
\begin{equation}
D_{m}= \frac{(M-m)!}{M!}\sum_{n=0}^M M^n  s(m,n) n
\label{DmStirling}
\end{equation}
and
\begin{equation}
E_m= \frac{(M-m)!}{M!}\sum_{n=0}^M M^n  s(m,n) n^2.
\label{EmStirling}
\end{equation}
Taking into account the dependence of $g^{(2)}(0)$ on the first and second moments of the photon number operator, cf. Eq. (\ref{g2definition}),  we can estimate $g^{(2)}(0)$ as
\begin{equation}
g_M^{(2)}(0)=\frac{\mu_2-\mu_1}{\mu_1^2}.
\end{equation}

Let us now address the effect  of losses. Under loss, the normally ordered moments of photon number operator are rescaled according to $\langle \hat{a}^{\dagger k}\hat{a}^k\rangle \rightarrow \eta^k \langle  \hat{a}^{\dagger k}\hat{a}^k\rangle $. Since the estimators $\mu_1$ and $\mu_2$ are designed to yield exact results for any state  containing no more than $M$ photons, the same rescaling applies to estimators of normally ordered moments. Consequently,  the estimators $\mu_1(\eta)$ and $\mu_2(\eta)$ which compensate for detection efficiency $\eta$ can be constructed from the estimators (\ref{DmStirling}) and (\ref{EmStirling}) for $\eta=1$ as follows
\begin{equation}
\mu_1(\eta)=\frac{\mu_1}{\eta},\qquad \mu_2(\eta) =\frac{\mu_2-\mu_1}{\eta^2}+\frac{\mu_1}{\eta}.
\end{equation}
Validity of these expressions can be confirmed by directly evaluating $D_m$ and $E_m$ for the lossy detector described by the response (\ref{Cmndefinition}) with $\eta<1$.
Consequently, ex-post compensation of losses during data processing does not change the value of the estimated parameter $g_M^{(2)}(0)$. This should not be confused with the effect of physical losses that occur before or during the detection. Such losses generally do influence the value of  $g_M^{(2)}(0)$, as discussed below.

In Fig.~\ref{figDE} we plot the coefficients $D_m$ and $E_m$ for two balanced multiplexed detectors with $10$ and $20$ channels, respectively. 
The coefficients $D_m$ and $E_m$ are all nonnegative and increase with $m$. The maximum value of $\mu_1$ and $\mu_2$ is given by $D_M$ and $E_M$, respectively.  For states with nonzero population of Fock states above $|M\rangle$ the estimators are biased and there will be a systematic difference between the true and estimated values of $\langle \hat{n}\rangle$ and $\langle \hat{n}^2\rangle$.  
\begin{figure}[b]
\includegraphics[width=\linewidth]{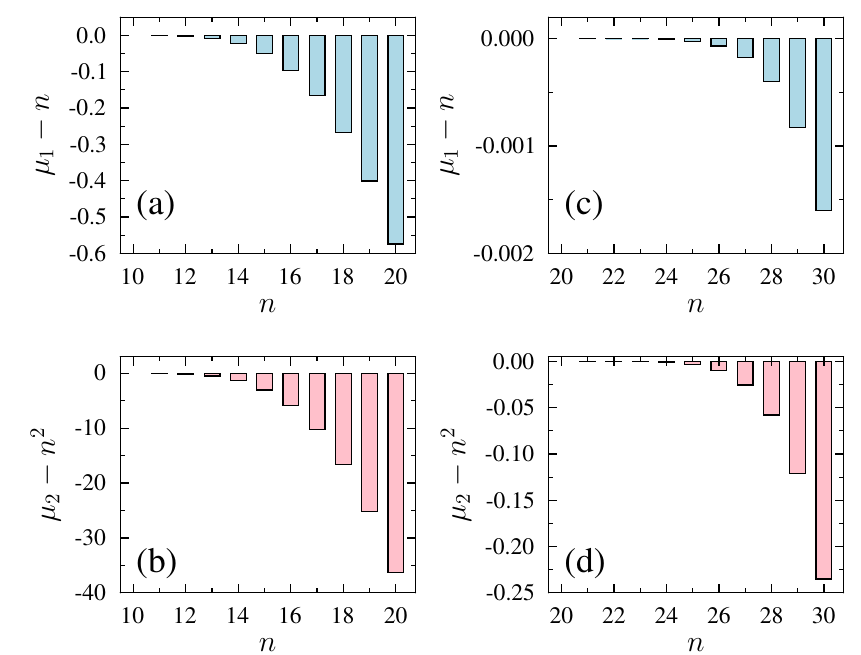}
\caption{Differences between the estimated and the true values of the first and second moments of the photon number operator are plotted for input Fock states $|n\rangle$ and two different numbers of detection channels $M=10$ (a,b) and $M=20$ (c,d). 
The other parameters read $\eta=1$ and $R_D=0$. Note the different scales on the horizontal axes for $M=10$ and $M=20$.}
\label{figmundif}
\end{figure}
Remarkably, the estimators can nevertheless perform very well even over a certain range of Fock states above $|M\rangle$ \cite{Fiurasek2025}. This is illustrated in Fig.~\ref{figmundif}, where we plot the differences $\mu_1-n$ and $\mu_2-n^2$ for input Fock states in the range $M+1$ to $M+10$. Interestingly, increasing the number of detection channels from $10$ to $20$ significantly  improves the performance of the estimators for these higher Fock states. For the $20$-channel detector the relative systematic error is of the order of $10^{-4}$ even for the Fock state $|30\rangle$.

\begin{figure}[t]
\includegraphics[width=\linewidth]{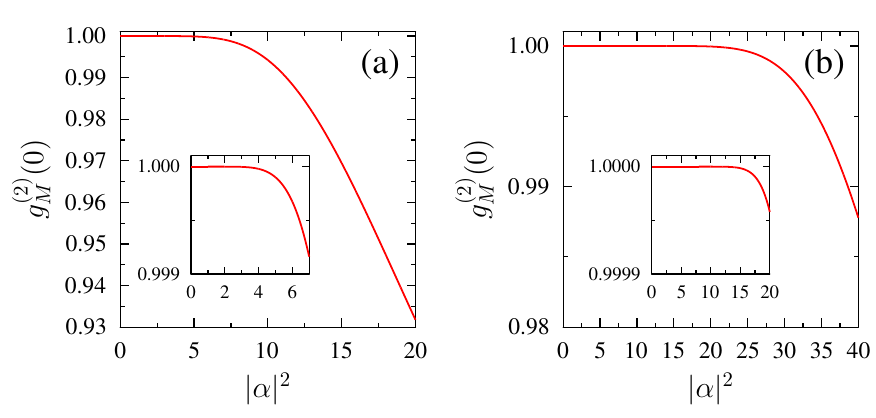}
\caption{Estimated second-order correlation function $g_M^{(2)}(0)$ for input coherent state $|\alpha\rangle$ is plotted as a function of the mean photon number $|\alpha|^2$ for two different numbers of detection channels $M=10$ (a) and $M=20$ (b). The insets show the behavior  of $g_M^{(2)}(0)$ in the region $|\alpha|^2<M$.} 
\label{figg2coherent}
\end{figure}

To benchmark the performance of the proposed estimation procedure we plot in Fig.~\ref{figg2coherent}  the estimate $g_M^{(2)}(0)$ for input coherent states with complex amplitude $\alpha$ and mean photon number $|\alpha|^2$. The click statistics of coherent states follows a binomial distribution, 
\begin{equation}
c_m={M \choose m } p^m (1-p)^{M-m},
\label{cmbinomial}
\end{equation}
where $p=1-\exp(-|\alpha|^2/M)$. As can be seen in Fig.~\ref{figg2coherent}, we obtain very precise estimates of $g^{(2)}(0)$ even for states with mean photon number significantly exceeding $1$. With the $20$-channel detector, the relative error of estimating $g^{(2)}(0)$ is of the order of $10^{-4}$ even for $|\alpha|^2=20$.

\begin{figure}[b!]
\includegraphics[width=\linewidth]{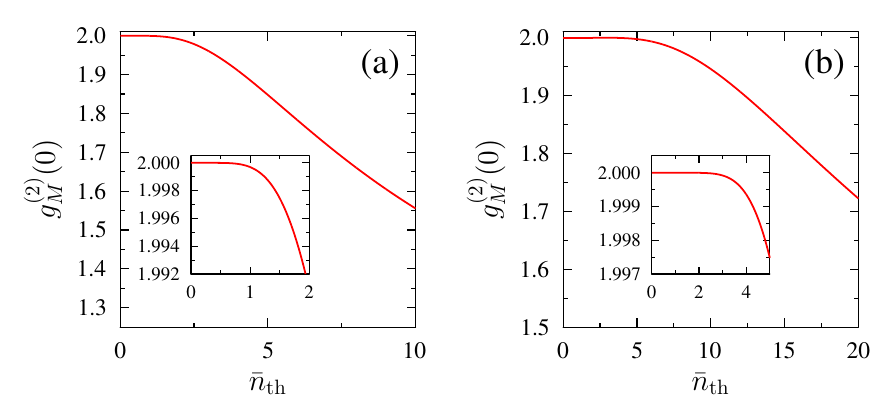}
\caption{Estimated second-order correlation function $g_M^{(2)}(0)$ for input  single-mode thermal  state is plotted as a function of the mean photon number $\bar{n}_{\mathrm{th}}$ for two different numbers of detection channels $M=10$ (a) and $M=20$ (b). } 
\label{figg2thermal}
\end{figure}

As a second example, consider thermal states that exhibit a super-Poissonian Bose-Einstein photon number distribution
\begin{equation}
p_n=\frac{1}{1+\bar{n}_{\mathrm{th}}} \left(\frac{\bar{n}_{\mathrm{th}}}{\bar{n}_{\mathrm{th}}+1}\right)^n.
\label{pnBose}
\end{equation}
For single-mode thermal light, $g^{(2)}(0)=2$ and does not depend on the mean photon number $\bar{n}_{\mathrm{th}}$. The click statistics corresponding to the photon number distribution (\ref{pnBose}) is given by the beta-binomial distribution \cite{Sperling2012a},
\begin{equation}
c_m= {M \choose m} \frac{M}{\bar{n}_{\mathrm{th}}}B\left(M+\frac{M}{\bar{n}_{\mathrm{th}}}-m,m+1\right),
\label{cmthermal}
\end{equation}
where $B(x,y)$ denotes the beta function. The Bose-Einstein photon number distribution has a heavier tail than the Poisson distribution, making contributions of large Fock states more important. This results in faster decrease of $g_M^{(2)}(0)$ with increasing $\bar{n}_{\mathrm{th}}$, as illustrated in Fig.~\ref{figg2thermal}. Nevertheless, observe that both plotted curves exhibit a plateau and the estimation is highly reliable for $\bar{n}\lesssim 1$ ($M=10$) and $\bar{n}\lesssim 4$ ($M=20$), respectively.

In  measurements with an ordinary HBT setup, it is well understood that the measurement yields a good approximation to the true value of $g^{(2)}(0)$ only when the input signal is sufficiently weak \cite{Stensson2018}. Since $g^{(2)}(0)$ does not depend on losses, it is possible to increase the detection range and suppress the systematic error of the  measurement  by introducing additional  losses with transmittance $\eta_D$ \cite{Stensson2018}. This technique can be directly applied also to  measurements with the $M$-channel  multiplexed  detector. One can repeatedly measure $g_M^{(2)}(0)$ for increasing losses (i.e., decreasing transmittance $\eta_D$) and stop the measurement once a plateau is reached.  Introducing losses is equivalent to moving from the right to the left along the $g_M^{(2)}(0)$  curves in Figs.~\ref{figg2coherent} and \ref{figg2thermal}. These figures illustrate that with the multiplexed detector we can reach the plateau and obtain  precise estimation of $g^{(2)}(0)$ even when the input signal is not really weak. 

The measurement of $g^{(2)}(0)$ is often utilized to characterize nonclassical states with sub-Poissonian photon number distribution. For such states, we can expect the estimation procedure to perform at least as well as for coherent states, because a sub-Poissonian photon number distribution is narrower than the Poissonian distribution with the same mean. As concrete examples, we consider here  Gaussian amplitude-squeezed coherent states and non-Gaussian single-photon-added coherent states.

\begin{figure}
\includegraphics[width=\linewidth]{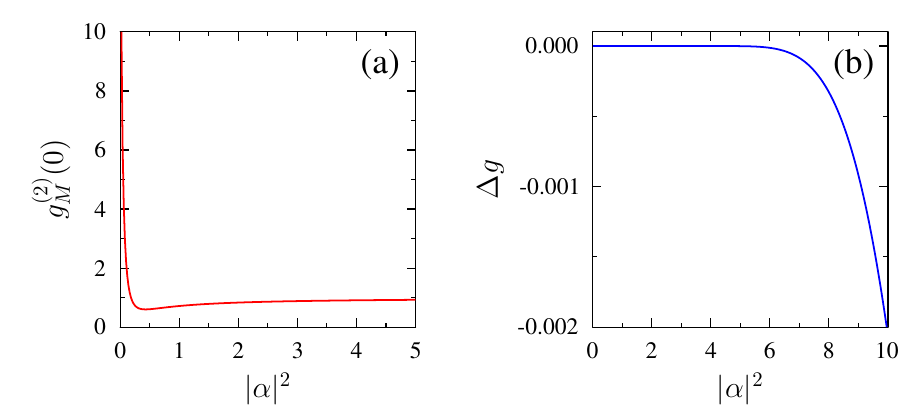}
\caption{Estimated second-order correlation function $g_M^{(2)}(0)$ for input  pure amplitude squeezed coherent state  is plotted as a function of $|\alpha|^2$ for $r=0.25$ and $M=10$ (a).
The difference between the estimated and the true values of the correlation function $\Delta g=g_M^{(2)}(0)-g^{(2)}(0)$ is plotted in panel (b).}
\label{figg2squeezed}
\end{figure}

The click statistics of a pure amplitude-squeezed coherent state with squeezing constant $r$ and coherent displacement $\alpha$ can be expressed as 
\begin{equation}
c_m={M \choose m} \sum_{j=0}^m (-1)^j {m \choose j} p_{0}\left(T_j\right),
\end{equation}
where $T_j=(M-m+j)/M$ and 
\begin{equation}
p_0(T)=\frac{2\exp\left[-\frac{2T|\alpha|^2}{2-T+Te^{-2r}}\right]}{\sqrt{(2-T+Te^{-2r})(2-T+Te^{2r})}}
\end{equation}
denotes the vacuum probability for the state transmitted via lossy channel with transmittance $T$.  In Fig. \ref{figg2squeezed}(a) we plot the estimated $g_M^{(2)}(0)$ as a function of $|\alpha|^2$ for $r=0.25$ and $M=10$.  Large values of $g_M^{(2)}(0)$ for small $|\alpha|^2$  reflect the behavior of squeezed vacuum that exhibits super-Poissonian statistics. As $|\alpha|^2$ increases,  $g_M^{(2)}(0)$ drops below $1$ and then begins to increase again since for very large $|\alpha|^2$ the coherent displacement becomes dominant.
Figure \ref{figg2squeezed}(b)  shows the difference between the estimated value $g_M^{(2)}(0)$ and the true value \cite{Grosse2007}
\begin{equation}
g^{(2)}(0)=1+\frac{\sinh^2(r)\cosh(2r)-2|\alpha|^2\sinh(r)e^{-r}}{[|\alpha|^2+\sinh^2(r)]^2}.
\end{equation}
We can see that, for the parameters considered, the difference between the estimated and true values is very small for $|\alpha|^2<10$.

\begin{figure}
\includegraphics[width=\linewidth]{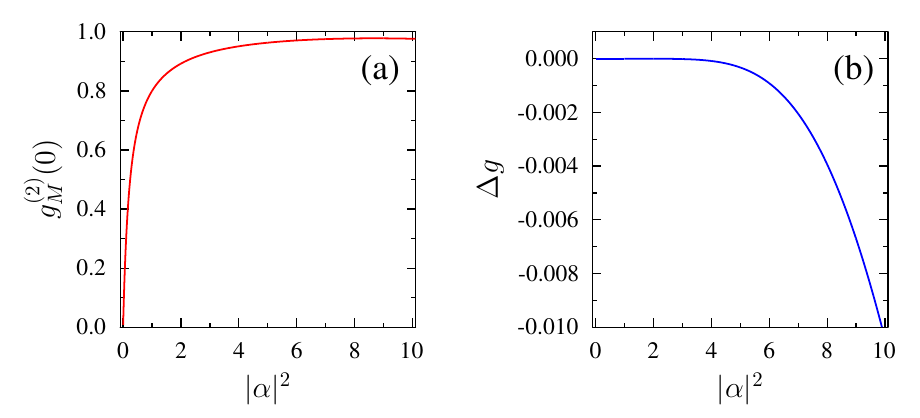}
\caption{The same as Fig.~6, but the results are shown for input single-photon-added coherent state $|\alpha,1\rangle$. }
\label{figg2added}
\end{figure}

We repeat this analysis also for a non-Gaussian  photon-added coherent state \cite{Agarwal1991,Zavatta2004,Barbieri2010,Kumar2013,Fadrny2024} defined as
\begin{equation}
|\alpha,1\rangle= \left(|\alpha|^2+1\right)^{-1/2} \hat{a}^{\dagger}|\alpha\rangle.
\end{equation}
The click statistics of the photon-added coherent state can be derived with the help of the following identity, which holds for any operator $\hat{\Pi}$,
\begin{equation}
\langle \alpha,1|\hat{\Pi}|\alpha,1\rangle=\frac{e^{-|\alpha|^2}}{|\alpha|^2+1} \frac{\partial^2}{\partial \alpha\partial\alpha^\ast} \left[e^{|\alpha|^2} \langle \alpha|\hat{\Pi}|\alpha\rangle\right].
\label{addedidentity}
\end{equation}
To obtain the click statistics of the photon-added coherent state we insert the binomial click statistics of coherent input states to the right-hand side of Eq.~(\ref{addedidentity}) and set 
\begin{equation}
\langle\alpha |\hat{\Pi}|\alpha\rangle={M \choose m}\left(1-x\right)^m x^{M-m},
\end{equation}
where $x=\exp(-|\alpha|^2/M)$. After some algebra, we obtain
\begin{eqnarray}
c_m&=&\frac{m}{M(1+|\alpha|^2)} {M \choose m} x^{M-m}(1-x)^{m-2} \nonumber \\
& & \times\left[1-x+\frac{|\alpha|^2}{M}(m-x)\right].
\end{eqnarray}
The results are plotted in Fig.~\ref{figg2added}. Once again  $g^{(2)}(0)$ can be determined with very good precision even for large displacements and we obtain the expected behavior showing the transition from the single-photon Fock state $|1\rangle$ at $\alpha=0$ to essentially coherent state at large $|\alpha|$ \cite{Zavatta2004}.

A few comments and warnings are in order. Since the photon number operator is unbounded, there may exist pathological states for which the proposed scheme may fail. Consider, for instance, a state which contains the Fock state $|N^2\rangle$   with probability $1/N$. By increasing $N$ the mean photon number of the state can become arbitrary large while the probability of $|N^2\rangle$ becomes arbitrarily small. 
The proposed procedure is applicable provided  that the photon number distribution behaves reasonably and  has a sufficiently rapidly decaying tail as $n$ increases. Such behavior can be potentially justified by considering the actual experimental scheme that generates the characterized signal. If no assumptions can be made
about the photon number distribution or its Fock-space cutoff,
 then the moments of the photon number operator cannot be reliably estimated from the click statistics \cite{Fiurasek2025}.

\section{$g_M^{(2)}(0)$ as nonclassicality witness}

To test the nonclassicality of the measured state,  we can introduce a nonclassicality criterion specifically designed for $g_M^{(2)}(0)$. Since $g_M^{(2)}(0)$ generally differs from the exact value $g^{(2)}(0)$, the nonclassicality criterion deviates from $g_M^{(2)}(0)<1$, especially for strong signals where saturation effects become significant. Specifically, this nonclassicality criterion reads 
\begin{equation}
g_M^{(2)}(0)< g_{M,\mathrm{c}}^{(2)}(0),
\label{g2criterion}
\end{equation}
where  $g_{M,\mathrm{c}}^{(2)}(0)$ is the value of $g_M^{(2)}(0)$ achieved by a coherent state with the same value of $\mu_1$ as the probed signal state. The inequality (\ref{g2criterion}) can be thus equivalently formulated  as  $\mu_2<\mu_{2,c}$,  where $\mu_{2,c}$ is the value of $\mu_2$   that we would obtain for a coherent state with the same $\mu_1$ as the measured state. 
To prove the criterion (\ref{g2criterion}), we show below that for input coherent states the dependence of $\mu_2$ on $\mu_1$ is a convex function.  As an example, we plot this dependence in Fig.~8 for $M=10$. The convexity of the curve implies that physically achievable points $(\mu_1 ,\mu_2)$ lying below this curve cannot be obtained as convex mixtures of the points on the boundary curve, i.e., they cannot be obtained by convex mixtures of coherent states. This is precisely captured by the inequality (\ref{g2criterion}). 

\begin{figure}[t!]
\includegraphics[width=\linewidth]{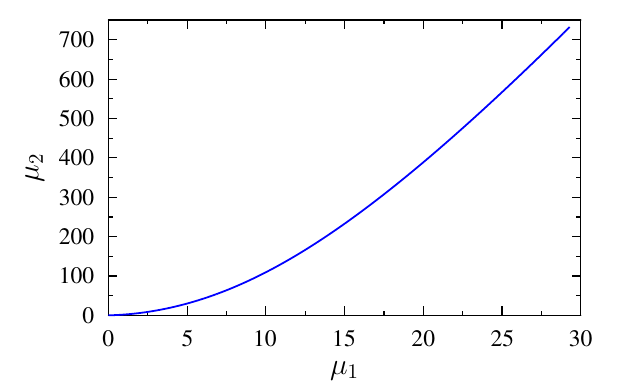}
\caption{Dependence of $\mu_2$ on $\mu_1$ for coherent-state inputs is plotted for a balanced multiplexed detector with $M=10$. } 
\label{xxx}
\end{figure}

For input coherent states, the dependence of $\mu_2$ on $\mu_1$  can be analytically expressed in a parametric form (see Appendix),
\begin{eqnarray}
\mu_{1,\mathrm{c}}&=&\sum_{k=1}^M\frac{M}{k }p^k ,
\label{mu1coh}
\end{eqnarray}
and
\begin{eqnarray}
\mu_{2,\mathrm{c}}&=&\mu_{1,c}+2M^2\sum_{k=2}^M \frac{H_{k-1}}{k}p^k.
\label{mu2coh}
\end{eqnarray}
Here  $p=1-\exp\left(-|\alpha|^2/M\right)$ is the click probability in the binomial distribution (\ref{cmbinomial}),  $H_k=\sum_{j=1}^k 1/j$ for $k\geq 1$ and we set $H_0=0$.
The convexity of the curve plotted in Fig.~8 can be proved by calculating its signed curvature. Since 
\begin{equation}
\frac{d \mu_{1,\mathrm{c}}}{d p}>0, \qquad 0<p<1,
\end{equation}
the convexity of the curve is equivalent to
\begin{equation}
K=\frac{d \mu_{1,\mathrm{c}}}{dp}\frac{d^2 \mu_{2,\mathrm{c}}}{d p^2}-\frac{d^2 \mu_{1,\mathrm{c}}}{dp^2}\frac{d \mu_{2,\mathrm{c}}}{d p} >0.
\label{Kinequality}
\end{equation}
Explicit calculation yields
\begin{equation}
K=M^3\sum_{j,k=0}^{M-1} p^{j+k-1}(j-k)(H_j-H_k).
\end{equation}
Since $H_j$ is strictly increasing with $j$, every summand is nonnegative, and at least one is positive for $M\geq2$. Consequently, $K>0$  for all $0<p<1$.

 Looking at Fig.~4 we can qualitatively state that for mean photon numbers below $M$ the criterion (\ref{g2criterion}) will very closely resemble the ordinary nonclassicality criterion $g^{(2)}(0)<1$. However, for stronger signals the saturation effects will start to play a role and the threshold on the right-hand side of Eq. (\ref{g2criterion}) will decrease well below 1.
The nonclassicality criterion (\ref{g2criterion}) complements the previously established nonclassicality criteria  based on matrices of moments \cite{Sperling2013,Sperling2015} or the criterion $Q_B<0$ based on the parameter 
\begin{equation}
Q_B=M\frac{\langle (\Delta c)^2\rangle}{\langle c\rangle (M-\langle c\rangle)}-1,
\label{QBcriterion}
\end{equation}
which detects sub-binomiality of the click statistics \cite{Sperling2012a}. In both criteria (\ref{g2criterion}) and $Q_B<0$ the threshold is determined by coherent-state input.
We would like to emphasize that there exist nonclassical states that are detectable via the criterion (\ref{g2criterion}) but do not exhibit sub-binomial click statistics. Note that general tight analytical nonclassicality criteria for click statistics of a perfectly balanced detector were derived recently, by relating the construction of the criterion to the Hausdorff moment problem \cite{Kovtoniuk2026}. Nevertheless, given the importance of the correlation function $g^{(2)}(0)$ in quantum optics, we find it relevant to formulate a nonclassicality criterion for the estimate $g_M^{(2)}(0)$. Moreover,  the criterion (\ref{g2criterion}) can be in principle extended also to unbalanced detectors, similarly to other specific criteria  \cite{Lee2016,Bohmann2019}.

\section{Unbalanced detector}
In practice, the detector may not be exactly balanced and the effective transmittance $T_j$ may be different for each detection channel. This may be either a consequence of imperfect alignment or a feature of the detector construction, e.g., when using a single fiber loop for temporal multiplexing \cite{Rehacek2003,Banaszek2003}. We parameterize the imbalances by factors $\epsilon_j$, defined as
\begin{equation}
T_j=\frac{\eta}{M}(1+\epsilon_j).
\end{equation}
We collect the coefficients $\epsilon_j$ into a vector $\vec{\epsilon}$.
By definition, we have
\begin{equation}
\sum_{j=1}^M \epsilon_j=0.
\end{equation}
 In practice, the factors $\epsilon_j$ can be determined from calibration measurements.

\begin{figure}[t!]
\includegraphics[width=\linewidth]{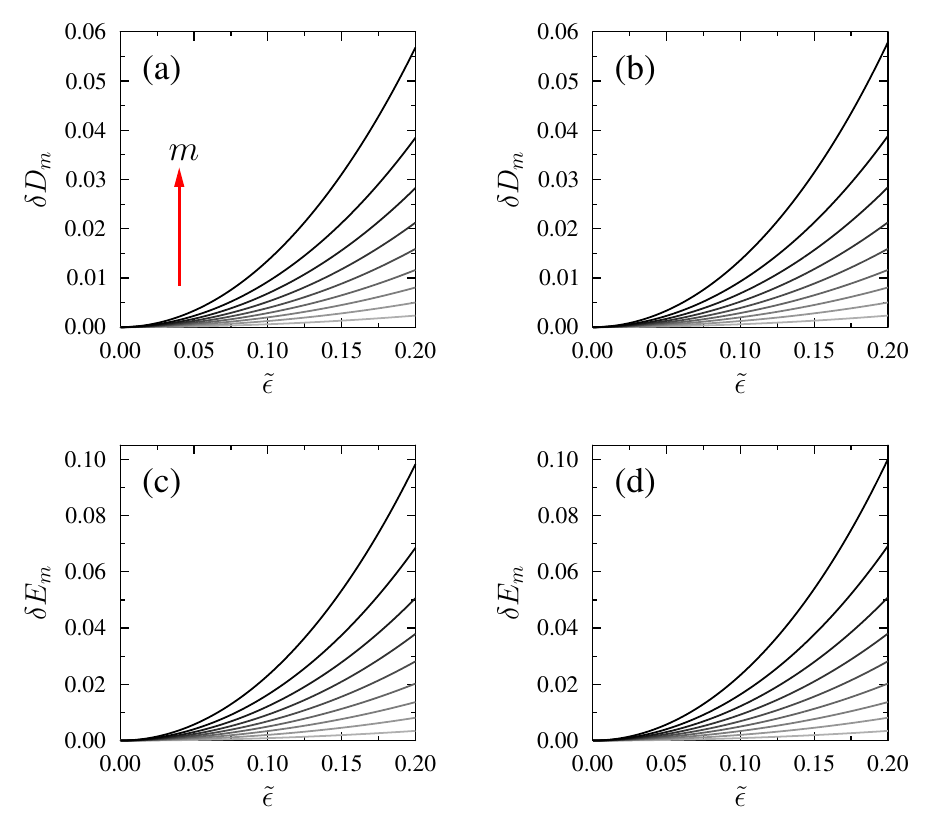}
\caption{The relative differences $\delta D_m$ and $\delta E_m$ are plotted as functions of the detector imbalance $\tilde{\epsilon}$ for the alternating shape of $T_j$ (a,c) and the linearly changing $T_j$ (b,d). Each curve on a graph corresponds to a specific $m$ and both $\delta D_m$ and $\delta E_m$  increase with $m$ for fixed $\tilde{\epsilon}$. The other parameters read $M=10$ and $\eta=1$.}  
\label{figunbalanced}
\end{figure}

An unbalanced detector offers, in principle, exponentially many distinguishable click patterns, since each combination of detector clicks and no-clicks may represent a specific outcome \cite{Rehacek2003,Fiurasek2024}. However, in practice it may be simplest to count just the total number of clicks and not distinguish all the detection patterns, especially when the imbalances $\epsilon_j$ are small. The click statistics of an unbalanced $M$-channel detector is still described by Eq. (\ref{cmsum}); however, the detector response $C_{mn}$ now depends on $\epsilon_j$.

 Importantly, if the $\epsilon_j$ are known, then $C_{mn}$ can be calculated efficiently, in polynomial time. In what follows we  neglect dark counts and set $\eta=1$.
Consider the input Fock state $|r\rangle$. Let  $\tilde{C}_{i,s,r}$ denote  the probability that all  $r$ input photons impinge on the first $i$ detectors and trigger clicks of $s$ detectors in total. We can establish the following recurrence relation for $\tilde{C}_{i,s,r}$,
\begin{equation}
\tilde{C}_{i+1,s,r}=\tilde{C}_{i,s,r}+\sum_{k=1}^{r-s+1} {r \choose k}\tilde{C}_{i,s-1,r-k} T_{i+1}^k.
\end{equation}
 Note that $\tilde{C}_{i,s,r}=0$ when $s>r$ or $i<s$. We have $\tilde{C}_{1,0,0}=1$ and  $\tilde{C}_{1,1,r}=T_1^r$ for $r\geq 1$,  which provides the starting values for the iteration. The response of the unbalanced detector is then given by
 \begin{equation}
 C_{mn}=\tilde{C}_{M,m,n}.
 \end{equation}
The coefficients $C_{mn}$ are symmetric functions of $\epsilon_j$, because the ordering of the detectors does not matter. 
This implies that 
\begin{eqnarray}
\left.\frac{\partial C_{mn}}{\partial \epsilon_j}\right|_{\vec{\epsilon}=0}&=&F_{mn}, \nonumber \\[2mm]
\left.\frac{\partial^2 C_{mn}}{\partial \epsilon_j \partial \epsilon_k}\right|_{\vec{\epsilon}=0}&=&G_{mn},  \quad j\neq k,  \\[2mm]
\left.\frac{\partial^2 C_{mn}}{\partial \epsilon_j ^2 }\right|_{\vec{\epsilon}=0}&=&J_{mn}, \nonumber
\end{eqnarray}
do not depend on $j$ and $k$. Consequently, the Taylor series expansion of $C_{mn}$  around $\vec{e}=0$ takes a simple from
\begin{equation}
C_{mn}(\epsilon)\approx  C_{mn}(0)+\frac{1}{2}(J_{mn}-G_{mn})\sum_{j} \epsilon_j^2 + O(\epsilon^3).
\end{equation}
It follows that to second order the changes in the response of the detector depend on a single quantity  
\begin{equation}
\tilde{ \epsilon} =\sqrt{ \frac{1}{M}\sum_{j=1}^M \epsilon_j^2}.
\end{equation}

As an illustration, we assume even $M$ and consider two shapes of $T_j$:  either odd-even alterations
\begin{equation}
T_j=\frac{1+(-1)^j \tilde{\epsilon}}{M},
\label{Tshape1}
\end{equation}
or a linear dependence 
\begin{equation}
T_j=\frac{1+x(2j-M-1)}{M},
\label{Tshape2}
\end{equation}
where $x=\sqrt{3/(M^2-1)}\tilde{\epsilon}$ ensures that the root-mean-square imbalance equals $\tilde{\epsilon}$. For both shapes (\ref{Tshape1}) and (\ref{Tshape2}) we have determined the estimators using the general matrix inversion formula (9) and calculated the relative differences 
\begin{equation}
\delta D_m=\frac{D_m(\vec{\epsilon})-D_m(0)}{D_m(0)}, \qquad \delta E_m=\frac{E_m(\vec{\epsilon})-E_m(0)}{E_m(0)}, 
\end{equation}
where $D_m(0)$ and $E_m(0)$ denote the estimators for balanced detectors. We plot $\delta D_m$ and $\delta E_m$ in Fig.~\ref{figunbalanced} for the alternating $T_j$ (a,c) and linearly varying $T_j$  (b,d). The results for the two considered shapes of $T_j$ are very similar, consistent with the fact that to second order the estimators depend on the single parameter $\tilde{\epsilon}$ only.
We have also repeated the analysis of the behavior of $g_M^{(2)}(0)$ performed in Section III for unbalanced detectors (with $\tilde{\epsilon}$ up to $0.2$). The conclusions remain unchanged and the results for unbalanced detectors closely match those for balanced detectors.

\section{Conclusions}

In the present paper we have provided a detailed theoretical analysis of a procedure for direct estimation of the correlation function $g^{(2)}(0)$ from click statistics. Our study reveals that a seemingly naive approach based on linear estimators can be very efficient and powerful and provide accurate estimate of $g^{(2)}(0)$ even for signals with rather large mean photon numbers.  The constructed estimators of photon number moments exhibit regular behavior and guarantee positivity of the estimated moments. On the other hand, the estimators become  biased for states with Fock-state populations above $|M\rangle$ and reach saturation for strong input signals. This can in principle be remedied by introducing additional controlled and tunable losses in the measurement setup. Importantly, it is not necessary to attenuate the signal to mean photon number values much smaller than $1$. The performance of the method improves with increasing number of channels. For comparison, we have considered detectors with $10$ and $20$ channels, which are currently experimentally achievable with spatial multiplexing \cite{Hlousek2019,Hlousek2024}.

We have proposed a nonclassicality criterion based on the  estimated $g_M^{(2)}(0)$ and rigorously proved it for balanced detectors. The photon number moments and correlation functions $g^{(k)}(0)$ can also be used to certify quantum non-Gaussianity of the input state  \cite{Hotter2026,Kalash2026,Racz2026}. Our work paves the way to applying these criteria to measurements with multiplexed binary detectors. Of particular interest is the criterion put forward in Ref. \cite{Hotter2026} and based on $g^{(2)}(0)$ and $g^{(3)}(0)$, because it is insensitive to losses and thus allows to certify quantum non-Gaussianity that is robust and survives any loss with nonzero transmittance.

While we have mainly considered balanced detectors, we have pointed out that the estimation procedure can be equally well applied also to unbalanced detectors. In practice, achieving exact balancing may be challenging and some small residual imbalances may remain present. If we can calibrate the effective transmittances of the detection channels, we can construct estimators tailored specifically to such an unbalanced detector. This may be particularly important when testing the nonclassicality of the detected states, to avoid falsely identifying them as nonclassical \cite{Lee2016}.

\begin{acknowledgments}
The author acknowledges support by the project OP JAC  CZ.02.01.01/00/23\_021/0008790  of the Ministry of Education, Youth, and Sports of the Czech Republic and EU.
The author used OpenAI's ChatGPT-5.6 Sol for grammar and style editing, assistance with selected calculations, and help in constructing the proof of the nonclassicality criterion.  All results obtained with the help of AI were checked by the author, who takes full responsibility for the final mathematical arguments, results, and conclusions.
\end{acknowledgments}

\section*{Data availability statement}
The data that support the findings of this article will be made openly available on Zenodo. 

\appendix*
\section{}
Here we derive the formulas  (\ref{mu1coh}) and (\ref{mu2coh}) for $\mu_{1,c}$ and $\mu_{2,c}$. We assume perfectly balanced detector with $\eta=1$ and $R_D=0$.
Our starting point is the generating function for the signed Stirling numbers of the first kind:
\begin{equation}
f_m (z):=z(z-1)\cdots(z-m+1)= \sum_{n=0}^{m}s(m,n)z^n,
 \label{fmdefinition}
\end{equation}
with $f_0(z)=1$.
Combining Eq. (\ref{fmdefinition}) with Eqs. (\ref{DmStirling}) and (\ref{EmStirling}), the coefficients $D_m$ and $E_m$ that define the estimators $\mu_1$ and $\mu_2$ can be written as
\begin{eqnarray}
 D_m&=&\left.\frac{(M-m)!}{M!} \left[z \frac{\partial}{\partial z} \right] f_m(z)\right|_{z=M}, \nonumber \\[3mm]
 E_m&=&\left.\frac{(M-m)!}{M!}\left[z \frac{\partial}{\partial z} \right] ^2 f_m(z)\right|_{z=M}.
 \label{eq:derivative-weights}
\end{eqnarray}
Upon combining these expressions with the binomial click statistics of coherent states (\ref{cmbinomial}), we obtain
\begin{eqnarray}
 \mu_{1,\mathrm c}(p)&=&\left. \left[z \frac{\partial}{\partial z} \right] G_M(z,p)\right|_{z=M}, \nonumber \\[2mm]
 \mu_{2,\mathrm c}(p)&=&\left. \left[z \frac{\partial}{\partial z} \right]^2G_M(z,p)\right|_{z=M},
 \label{mu12generators}
\end{eqnarray}
where 
\begin{equation}
 G_M(z,p)=\sum_{m=0}^{M}{z\choose m}p^m(1-p)^{M-m}.
 \label{Gfunction}
\end{equation}
To proceed, we formally extend the upper limit of the summation in Eq. (\ref{Gfunction}) to infinity.  The generalized binomial theorem then gives the formal identity
\begin{equation}
 (1-p)^M\sum_{m=0}^{\infty}{z\choose m}
       \left(\frac{p}{1-p}\right)^m
       =(1-p)^{M-z}.
 \label{eq:binomial}
\end{equation}

For a power series in $p$, we write $[\cdot]_{\leq M}$ for truncation after its $p^M$ term. Consequently, function $G_M(z,p)$ satisfies the exact polynomial identity
\begin{equation}
 G_M(z,p)=[(1-p)^{M-z}]_{\leq M}.
 \label{GMtruncation}
\end{equation}
 Differentiation with respect to $z$ commutes with this finite truncation. 
 
 On inserting the expression (\ref{GMtruncation}) into Eq. (\ref{mu12generators})
we obtain 
\begin{eqnarray}
 \mu_{1,\mathrm c}&=&M[L(p)]_{\leq M},\nonumber \\[2mm]
 \mu_{2,\mathrm c}&=&M[L(p)]_{\leq M}+M^2[L(p)^2]_{\leq M},
 \label{eq:truncated-log}
\end{eqnarray}
where $L(p)=-\ln(1-p)$. Since
\begin{eqnarray}
 L(p)&=&\sum_{k=1}^{\infty}\frac{p^k}{k}, \nonumber \\[2mm]
 L^2(p) &=&\sum_{k=2}^\infty \sum_{a=1}^{k-1}\frac{p^k}{a(k-a)} 
 =\sum_{k=2}^\infty\frac{2H_{k-1}}{k} p^k,
 \label{eq:harmonic-identity}
\end{eqnarray}
Eq.~(\ref{eq:truncated-log}) proves, for every $M\geq1$,
\begin{eqnarray}
 \mu_{1,\mathrm c}(p)&=&M\sum_{k=1}^{M}\frac{p^k}{k}, \nonumber \\[2mm]
 \mu_{2,\mathrm c}(p)&=&\mu_{1,\mathrm c}(p)   +2M^2\sum_{k=2}^{M}\frac{H_{k-1}}{k}p^k.
\end{eqnarray}

\end{document}